\documentclass[12pt,a4paper]{article}
\pdfoutput=1
\usepackage{graphicx,hyperref,wrapfig,cite,color,subfigure,amssymb,amsmath,epsfig,
inputenc}
\hypersetup{
     colorlinks=true,       % false: boxed links; true: colored links
     linkcolor=red,          % color of internal links
     citecolor=green,        % color of links to bibliography
     filecolor=magenta,      % color of file links
     urlcolor=blue           % color of external links
}
\newcommand{\newc}{\newcommand}
\def\u#1{\verb!#1!\endgroup}

\newc{\HW}{\textsf{HERWIG}}
\newc{\TAUOLA}{\textsf{TAUOLA}}
\newc{\ThePEG}{\textsf{ThePEG}}
\newc{\HWPP}{\textsf{Herwig++}}
\newc{\evt}{\textsf{EvtGen}}
\newc{\fortran}{\textsf{FORTRAN}}
\newc{\decayer}{\textsf{Decayer}}

\newc{\HWPPClass}[1]{\href{http://herwig.hepforge.org/doxygen/classHerwig_1_1#1.html}{\textsf{#1}}}
\newc{\ThePEGClass}[1]{\href{http://thepeg.hepforge.org/doxygen/classThePEG_1_1#1.html}{\textsf{#1}}}
\newc{\HWPPParameter}[2]{\href{http://herwig.hepforge.org/doxygen/#1Interfaces.html\##2}{{\bf #2}}}
\newc{\ThePEGParameter}[2]{\href{http://thepeg.hepforge.org/doxygen/#1Interfaces.html\##2}{{\bf #2}}}
\newc{\HWPPParameterValue}[3]{\href{http://herwig.hepforge.org/doxygen/#1Interfaces.html\##2}{{\bf [#2=#3]}}}
\newc{\HWPPParameterValueB}[3]{\href{http://herwig.hepforge.org/doxygen/#1Interfaces.html\##2}{{\bf [#3]}}}
\newc{\ThePEGParameterValue}[3]{\href{http://thepeg.hepforge.org/doxygen/#1Interfaces.html\##2}{{\bf [#2=#3]}}}

\begin{document}
\tolerance=100000
\thispagestyle{empty}
\setcounter{page}{0}
 \begin{flushright}
IPPP/13/88\\
MCnet-13-15\\
DCPT/13/176\\
DESY 13-186\\
KA-TP-31-2013\\
ZU-TH 23/13

\end{flushright}
\begin{center}
{\Large \bf Herwig++ 2.7 Release Note}\\[0.7cm]

J.~Bellm$^1$,
S.~Gieseke$^1$,
D.~Grellscheid$^2$,
A.~Papaefstathiou$^3$,
S.~Pl\"atzer$^4$,
P.~Richardson$^{2}$,
C.~R\"ohr$^1$,
T.~Schuh$^1$,
M.~H.~Seymour$^{5}$,
A.~Si\'odmok$^5$,
A.~Wilcock$^2$,
B.~Zimmermann$^1$

E-mail: {\tt herwig@projects.hepforge.org}\\[1cm]

$^1$\it Institut f\"ur Theoretische Physik, Karlsruhe Institute of Technology.\\[0.4mm]
$^2$\it IPPP, Department of Physics, Durham University.\\[0.4mm]
$^3$\it Institut f\"ur Theoretische Physik, University of Z\"urich.\\[0.4mm]
$^4$\it Theory Group, DESY Hamburg.\\[0.4mm]
$^5$\it School of Physics and Astronomy, University of Manchester.
\end{center}

\vspace*{\fill}

\begin{abstract}{\small\noindent A new release of the Monte Carlo
    event generator \HWPP\ (version 2.7) is now available. This version comes with
    a number of improvements including: 
    an interface to the Universal FeynRules Output~(UFO) format allowing the
    simulation of a wide range of new-physics models;
    developments of the Matchbox framework for next-to-leading order~(NLO) simulations;
    better treatment of QCD radiation in heavy particle decays in new-physics models;
    a new tune of underlying event and colour connection parameters that allows a good simultaneous description of both Tevatron and LHC underlying event data and the double-parton scattering parameter $\sigma_{\!\textit{eff}}$.
  }
\end{abstract}

\tableofcontents
\setcounter{page}{1}

\section{Introduction}

The last major public version (2.6) of \HWPP\ is described in great
detail in \cite{Bahr:2008pv,UpdatedManual,Bahr:2008tx,Bahr:2008tf,
  Arnold:2012fq}. This release note describes all changes since
version 2.6.  The manual will be updated to reflect these changes and
this release note is only intended to highlight these new features and
the other minor changes made since version~2.6.

Please refer to \cite{Bahr:2008pv} and the present paper if using
version 2.7 of the program.

\subsection{Availability}

The new program version, together  with other useful files and information,
can be obtained from the following web site:
\begin{quote}\tt
       \href{http://herwig.hepforge.org/}{http://herwig.hepforge.org/}
\end{quote}
  In order to improve our response to user queries, all problems and requests for
  user support should be reported via the bug tracker on our wiki. Requests for an
  account to submit tickets and modify the wiki should be sent to 
  {\tt herwig@projects.hepforge.org}.

  \HWPP\ is released under the GNU General Public License (GPL) version 2 and 
  the MCnet guidelines for the distribution and usage of event generator software
  in an academic setting, which are distributed together with the source, and can also
  be obtained from
\begin{quote}\tt
 \href{http://www.montecarlonet.org/index.php?p=Publications/Guidelines}{http://www.montecarlonet.org/index.php?p=Publications/Guidelines}
\end{quote}

\section{Beyond the Standard Model Simulations}

\subsection{Universal FeynRules Output}

  In order to avoid having to explicitly code the Feynman rules for new-physics
models, \HWPP\ now includes an interface to programs which automatically calculates Feynman rules via
the Universal FeynRules Output~(UFO) format~\cite{Degrande:2011ua}.
This format can be generated by a number of 
programs, such as FeynRules~\cite{Christensen:2008py,Alloul:2013bka} and Sarah~\cite{Staub:2008uz,Staub:2012pb,Staub:2013tta}.

The directory containing the UFO output can automatically be converted to generate the code
required by \HWPP\ using
\begin{quote}\tt
ufo2herwig NAME\_OF\_UFO\_DIRECTORY
\end{quote}
which will automatically generate both code and input files. The code can be compiled
using the automatically generated {\tt Makefile} and events simulated using the automatically
generated {\tt LHC-FRModel.in} input file.

We expect that following this release there will be no further 
hard-coded new-physics models added to \HWPP\ and that future models
will be included using the UFO interface.

\subsection{Other Changes}

We have added the R-parity violating supersymmetric model including the option of
both bi- and tri-linear R-parity violating couplings, although not both simultaneously.
In addition later releases in the 2.6 series had already included 
Little Higgs models with and without the conservation of T-parity.

The machinery for the automatic generation of particle decays in new-physics
 models has been restructured to
allow the simulation of higher-multiplicity decays. So far only
the 4-body decays of scalar bosons to fermions, which are important for the decays of the Higgs boson,
the lightest top squark in supersymmetric models and the lightest tau slepton in R-parity
violating models, are included. However, the infrastructure will allow the inclusion of further 4-body
and higher-multiplicity decays in the future if required.

\section{NLO Development}

\subsection{Decay Chains}

Previous versions of \HWPP\ have included next-to-leading (NLO) 
POsitive Weight Hardest Emission Generator~(POWHEG)
scheme corrections\cite{Nason:2004rx,Frixione:2007vw} to a number of
processes.  In this release additional POWHEG-style corrections are
included for the Standard Model decay $t \rightarrow Wb$ and a range
of Beyond the Standard Model~(BSM) decays.  These corrections allow
the highest $p_T$ emission in the parton shower to be generated using
the real-emission matrix element.  However, we have not
implemented the $\bar{B}$ function that would be required to simulate
the decays with full NLO accuracy.  The correction has been documented
in detail in\cite{Richardson:2013jca} together with examples
illustrating the impact of the correction on a number of decays.

\subsubsection{Top Quark Decay}
Simulation of the hardest emission in the decay $t \rightarrow Wb$ is
implemented, using the POWHEG scheme, in the \HWPPClass{SMTopPOWHEGDecayer}
class.  The singular regions of the real-emission matrix element are
separated using the massive Catani-Seymour dipole formalism
in\cite{Catani:2002hc}.  The implementation was validated in
\cite{Richardson:2013jca} by comparing distributions generated using
the POWHEG-style correction with those simulated with the existing
\HWPP\ hard and soft matrix element corrections.  Whilst the two
approaches were found to give very similar results, the matrix element
corrections consistently produced slightly softer distributions.  This
difference is due to the soft matrix element correction being applied
to multiple hard emissions in the shower, whereas the POWHEG
correction applies only to the hardest.

\subsubsection{Decays of BSM Particles}
The POWHEG-style correction for two-body BSM decays is implemented in
the \HWPPClass{GeneralTwoBodyDecayer} class.  The real-emission matrix
element is calculated in the decayer class of the leading-order decay
using internal helicity-amplitude code.  The Catani-Seymour dipole
formalism\cite{Catani:1996vz, Catani:2002hc} is used to describe the
singular behaviour of the real-emission matrix element.  In this
approach, dipoles describing quasi-collinear radiation from massive
vector bosons are not well defined.  Therefore, fermion to fermion
vector, scalar to scalar vector and tensor to vector vector decays are
limited to cases in which final-state coloured vector particles are
massless.  The vector to fermion fermion and vector to scalar scalar
decays do, however, include radiation from massive incoming vector
particles.  Decays are performed in the rest frame of the decaying
particle and therefore the dipole describing the singular behaviour of
this particle will only contain a universal soft contribution.  This
is a well defined, spin-independent function.

Table~\ref{tab:spins} shows the combinations of incoming and outgoing
spins for which the POWHEG-style correction is included.  Each spin
structure is implemented for the colour flows given in
Table~\ref{tab:colour}. However, scenarios with coloured tensor
particles have not been considered and, therefore, decays involving
incoming tensor particles are limited to colour flows in which the
tensor is a colour singlet.

\renewcommand{\arraystretch}{1.05}
\begin{table}
\parbox{.45\linewidth}{
\begin{center}
\small
\begin{tabular}{|c|c|}
  \hline
  \textbf{Incoming} & \textbf{Outgoing} \\   \hline
  Scalar & Scalar Scalar \\ \hline
  Scalar & Scalar Vector* \\ \hline
  Scalar & Fermion Fermion \\ \hline
  Fermion & Fermion Scalar \\ \hline
  Fermion & Fermion Vector* \\ \hline
  Vector & Scalar Scalar \\ \hline
  Vector & Fermion Fermion \\ \hline
  Tensor & Fermion Fermion \\ \hline
  Tensor & Vector Vector* \\ \hline
\end{tabular}
\end{center}
\caption{Spin combinations for which the POWHEG-style correction has
  been applied.  Corrections to the decays marked * are not included
  for massive, coloured vector \mbox{particles.} }
\label{tab:spins}
}
%\end{table}
\hfill
\parbox{.45\linewidth}{
\renewcommand{\arraystretch}{1.305}
%\begin{table}
\begin{center}
\small
\begin{tabular}{|c|c|}
  \hline
  \textbf{Incoming} & \textbf{Outgoing} \\ \hline
  0 & 3 \(\bar{3}^{\dag}\) \\ \hline
  0 & 8 \(8^{ \dag}\) \\ \hline
  3 & 3 0 \\ \hline
  \(\bar{3}\) & \(\bar{3}\) 0 \\ \hline
  3 & 3 8 \\ \hline
  \(\bar{3}\) & \(\bar{3}\) 8 \\ \hline
  8 & 3 \(\bar{3}\) \\ \hline
\end{tabular}
\end{center}
\caption{Colour flows for which the POWHEG-style correction has been
  applied. For tensor particles, corrections are only included for
  colour flows marked \(^{\dag}\).\vspace{6.5mm}}
\label{tab:colour}
}
\end{table}

The POWHEG-style correction in the \HWPPClass{GeneralTwoBodyDecayer} class
may also be used to generate the hardest final-state emission in $2
\rightarrow 2$ processes that proceed via a colourless
\mbox{$s$-channel} resonance\footnote{For example, $q \bar{q}
  \rightarrow G \rightarrow q \bar{q}$, where $G$ is the lightest
  Randall-Sundrum graviton.}.  This is done by simulating the decay of
the intermediate particle, including the emission from the POWHEG-style
correction, and replacing the final-state particles in the hard
$2 \rightarrow 2$ process with those generated in the decay.  This
procedure is performed in the \HWPPClass{Evolver} class and will
automatically be attempted for all suitable $2 \rightarrow 2$
processes when POWHEG corrections are switched on.

\subsection{Matchbox 2.0\boldmath{$\beta$}}

Several improvements and enhancements have been developed for the
Matchbox NLO framework, of which we introduce version 2.0$\beta$ along
with this release.

In particular, the fixed-order part has been extensively tested and
pushed to calculating state-of-the-art $2\to 4$ processes at hadron
colliders. The interfaces to implement new processes have been
significantly simplified, and implementations of various scale choices
are now available in a much more transparent way. A generic interface
to matrix elements provided via the Binoth Les Houches Accord version
2 \cite{Alioli:2013nda} is now included. Matchbox also provides
facilities to implement subtraction algorithms different to the
default Catani-Seymour dipole \cite{Catani:1996vz} implementation
\cite{Platzer:2011bc}.

The matching structures have been rewritten completely to reflect
recent developments on the structure of consistent NLO matching and
the various ways to assign uncertainties to be described in detail
elsewhere \cite{Platzer:Matching}. Currently, matching to the dipole
shower is fully supported and a preliminary version of POWHEG-type
matching is available as well. Since the matching subtractions are
handled in a very generic way, we foresee matching to the default
angular ordered shower in one of the forthcoming versions.

A number of processes at $e^+e^-$, $ep$ and $pp$ colliders are
included at NLO, amongst them $pp\to V(+jet)$, where $V=W,Z$ including
the decay to lepton pairs, and $gg\to h$ and $b\bar{b}\to h$ inclusive
Higgs production. Electroweak Higgs plus two and three jet production
at NLO, as recently presented in \cite{Campanario:2013fsa}, is based on
the Matchbox framework and will
become publicly available soon.

\section{Higgs boson pair production}

\begin{figure}[!htb]
    \includegraphics[width=0.5\linewidth]{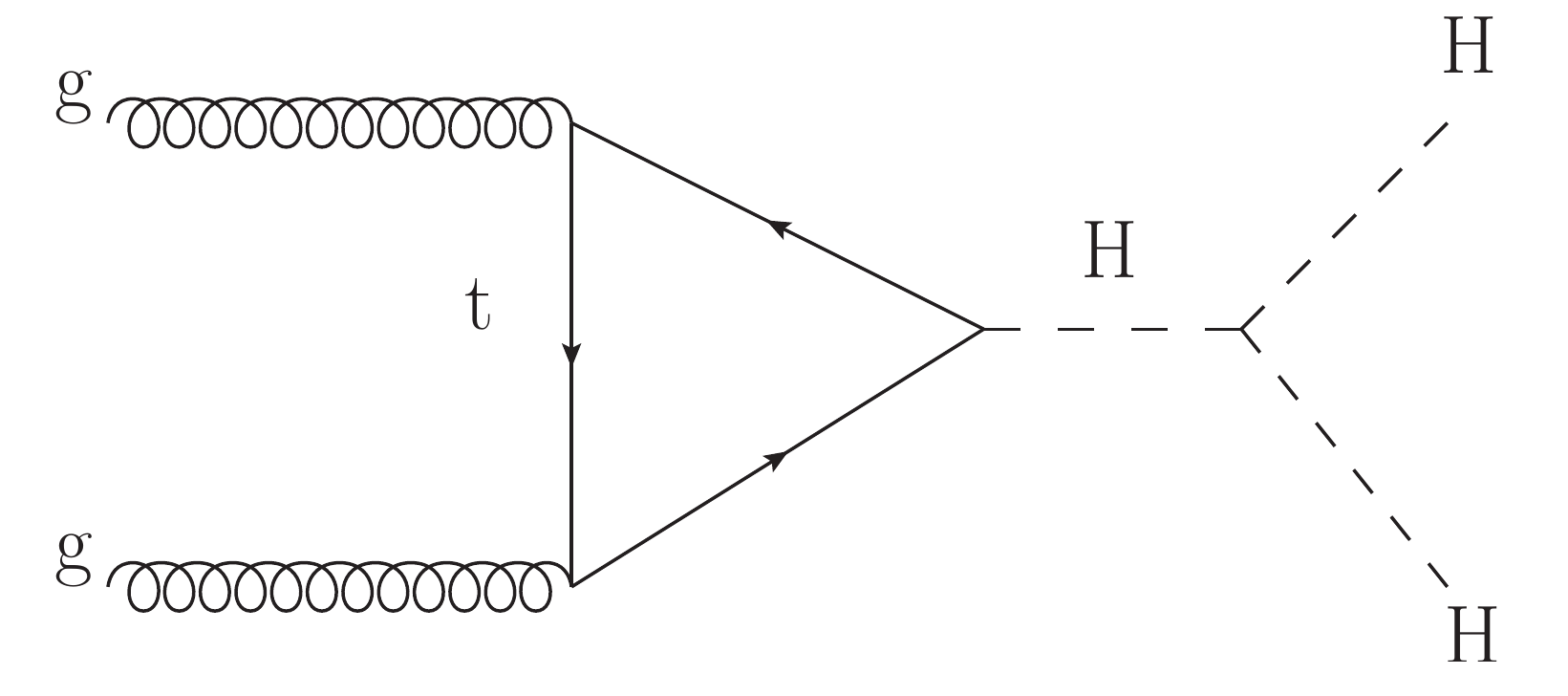}
    \includegraphics[width=0.5\linewidth]{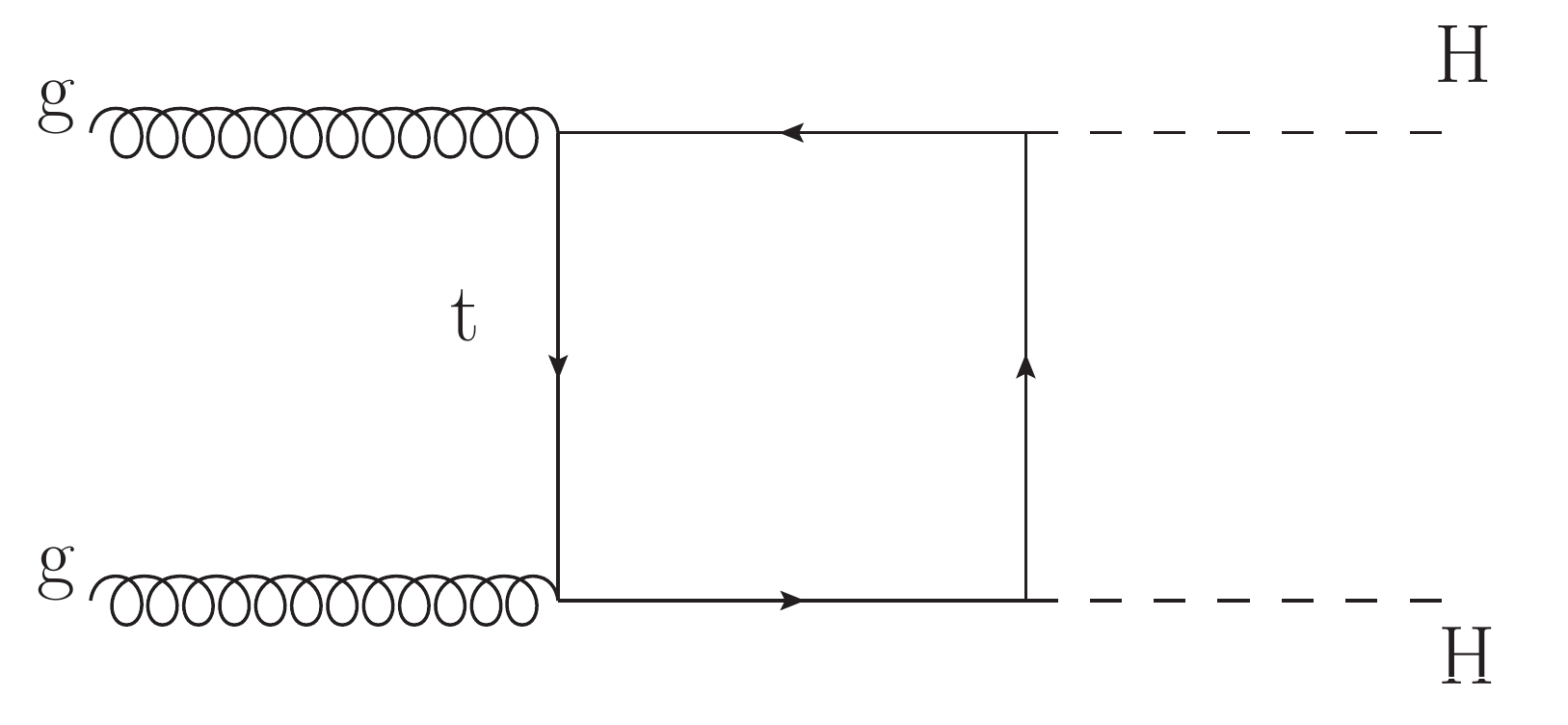}
  \caption{The Higgs pair production diagrams contributing to the
    gluon fusion process at LO are shown.}
  \label{fig:HHdiagrams}
\end{figure}

Higgs boson ($H$) pair production in the Standard Model (SM) at hadron
colliders is dominated by the gluon fusion initial
states~\cite{Plehn:1996wb, Dawson:1998py, Baur:2002qd, Baur:2003gp,
  Dolan:2012rv, Papaefstathiou:2012qe, Goertz:2013kp, Goertz:2013eka,
  Barr:2013tda, Dolan:2013rja, deFlorian:2013jea}.\footnote{Other modes, like
  $qq \rightarrow qqHH$,$VHH$, $t \bar{t} HH$ are a factor of 10-30
  smaller~\cite{Djouadi:1999rca,Gianotti:2002xx, Moretti:2004wa,
    Baglio:2012np}.} At leading order, the process is loop-initiated,
and the dominant diagrams contributing towards it are shown in
Fig.~\ref{fig:HHdiagrams}, where the loops can contain either a top
quark or a bottom quark.

 The triangle diagram can only contain initial-state gluons in a
 spin-0 state, whereas the box contribution can contain both spin-0
 and spin-2 configurations. Therefore, there are two Lorentz
 structures involved in the box diagram matrix element. At LO, we may
 write, schematically
\begin{equation}\label{eq:sigmaHHschem}
\sigma_{HH}^{LO} = \left| \sum_q ( \alpha_q C^{(1)}_{q,\mathrm{tri}} + \beta_q C^{(1)}_{q,\mathrm{box}})
\right|^2 +\left| \sum_q  \gamma_q  C^{(2)}_{q,\mathrm{box}} \right|^2 \;,
\end{equation}
where $C_{q,\mathrm{tri}}^{(1)}$ represents the matrix element for the
triangle contributions and $C_{q,\mathrm{box}}^{(i)}$ represents the
matrix element for the two Lorentz structures ($i = 1,2$) coming from
the box contributions~\cite{Glover:1987nx, Plehn:1996wb}, for each of
the quark flavours $q = \{t,b\}$.

The parameters $\alpha_q$, $\beta_q$ and $\gamma_q$ for quark flavour
$q$ are given in terms of the Standard Model Lagrangian parameters by
\begin{eqnarray}\label{eq:sigmadefs}
\alpha_q = \lambda y_q\;, \nonumber \\
\beta_q = \gamma_q = y_q^2 \;,
\end{eqnarray}
where $q =\{ t,b\}$. The dimensionless (and normalised) triple Higgs
coupling $\lambda$ is defined by
\begin{equation}\label{eq:lambda}
G_{HHH} = \lambda \frac{3 M_H^2}{v}\;,
\end{equation}
where $v \simeq 246$~GeV is the Higgs vacuum expectation value.
$y_q$ is the $Hq\bar{q}$ coupling (as defined after
electroweak symmetry breaking and assumed to be real), normalised to
the SM value, defined by
\begin{equation}
G_{HQ} = y_q \frac{M_q}{v}\;,\;\; (Q = \{B,T\})\;.
\end{equation}

The implementation of the process in the \HWPP\ event
generator was performed using the available \textsf{HPAIR}
code~\cite{hpair}. The model is present in
\texttt{Contrib/HiggsPair}.

The following options currently exist for the Higgs pair production
process:
\begin{itemize}
\item{\texttt{Process}: \texttt{All}, \texttt{ggToTriangleTohh},
    \texttt{ggToBoxTohh}, \texttt{ggToHToh}, where, respectively, they
    correspond to: `all SM $gg\rightarrow hh$ subprocesses', `only SM $gg\rightarrow hh$
    triangle subprocess', `only SM $gg\rightarrow hh$ box subprocess', ` all
    $gg\rightarrow hh$ subprocess, with heavy Higgs'. The heavy Higgs mass should
    be set in the input file with PDG id \texttt{35}.}
\item{\texttt{SelfCoupling}: set to the value of $\lambda$ as defined
    by Eq.~\ref{eq:lambda}.}
\item{\texttt{hhHCoupling}: set to the value of the (light Higgs-light
    Higgs-heavy Higgs) coupling.}
\end{itemize}

Validation of the implementation has been performed using an
equivalent \textsf{MadGraph} model~\cite{Alwall:2011uj, rikkert}, implemented in a
similar (but independent) way using functions taken from
\textsf{HPAIR}. We have made comparisons of several distributions and
the total cross section between the two implementations. We have also
confirmed that the total cross section output obtained by
using \HWPP\ matches that obtained using \textsf{HPAIR} at
leading order for various PDF sets and the scale choice
$\sqrt{\hat{s}}$.\footnote{\textsf{HPAIR} was modified to use the
  \textsf{LHAPDF} library~\cite{joselilin}.} 

\section{Underlying Event Tuning}
Multi-parton interaction (MPI) models are very successful in describing
soft inclusive and underlying event data over a wide range of centre-of-mass
energies. However, it has long been thought that they could not do this
while simultaneously describing the direct measurements of double parton
scattering. The latter are parametrised by a cross section normalization
factor, $\sigma_{\!\textit{eff}}$, which is related to the geometric size of
the colliding hadrons. The experimental
data\cite{Abe:1997xk,Abazov:2009gc} are consistent with
$\sigma_{\!\textit{eff}}\approx14\,\mathrm{mb}$\cite{Bahr:2013gkj},
while tunes to underlying event data typically predict
$\sigma_{\!\textit{eff}}=20-40\,\mathrm{mb}$\cite{Abramowicz:2013iva}.

This issue was considered in Ref.~\cite{Seymour:2013qka}. It was
shown that a good description of both underlying event and double parton
scattering data \emph{can} be obtained if one includes the latter in the
data being fit to with a sufficiently high weight. The point is that
while the fit to the underlying event data has a minimum at
$\sigma_{\!\textit{eff}}\approx21\,\mathrm{mb}$, this is the minimum of
a long thin valley in parameter space and reasonable fits can be
obtained up to significantly higher $\sigma_{\!\textit{eff}}$
values. The final result of Ref.~\cite{Seymour:2013qka} has
$\sigma_{\!\textit{eff}}\approx15\,\mathrm{mb}$ and gives a good
description of the underlying event data from Tevatron's lowest energy
point\cite{CDFUEscan}, $\sqrt{s}=300\,\mathrm{GeV}$ to the LHC's
highest\cite{Aad:2010fh}, $\sqrt{s}=7\,\mathrm{TeV}$.

\HWPP\ version 2.7 is released together with the tune of
Ref.~\cite{Seymour:2013qka}, \textsc{ue-ee-5-mrst}, by default. Other
related tunes can be obtained from
\href{https://herwig.hepforge.org/trac/wiki/MB_UE_tunes}{the \HWPP\ tunes page}.

\section{Uncertainties in Shower Algorithms}

Evaluating uncertainties within parton-shower algorithms is becoming
more and more important. Though no fully detailed accord is yet
available, we introduce some first measures for scale variations using
both parton-shower modules.

\subsection{Angular Ordered Shower}

Within the angular-ordered shower, a variation of the argument of the
strong coupling can be set using
\begin{quote}\tt
set /Herwig/Shower/AlphaQCD:RenormalizationScaleFactor x
\end{quote}
which will result in evaluating $\alpha_s(x q_\perp^2)$ for each
shower emission, where $q_\perp^2$ is the dynamic scale chosen by the
shower for each emission. We note that such variations are being
discussed as typically overestimating the uncertainty, and improved
schemes may be provided in the future.

A variation of the hard veto scale (as relevant to processes with jets
at the level of the hard process), can be achieved by using
\begin{quote}\tt
set /Herwig/Shower/Evolver:HardScaleFactor x
\end{quote}
which will scale the $p_\perp$ veto value for shower emissions by the
factor $x$.

\subsection{Dipole Shower}

The dipole shower implementation naturally supports variations of all the
scales involved, {\it i.e.} renormalization, factorization and hard
shower scales (similar to a resummation scale). The scale factors can
be set through the following interfaces
\begin{quote}\tt
cd /Herwig/DipoleShower

set DipoleShowerHandler:RenormalizationScaleFactor x

set DipoleShowerHandler:FactorizationScaleFactor x

set DipoleShowerHandler:HardScaleFactor x
\end{quote}
The hard scale factor setting is ignored, when running in MC@NLO mode,
and the corresponding value is obtained from the matching object in
place to ensure consistency. The corresponding value in the
matching object can be set by
\begin{quote}\tt
set /Herwig/MatrixElements/Matchbox/DipoleMatching:HardScaleFactor x
\end{quote}

\section{Other Changes}

A number of other more minor changes have been made.
The following changes have been made to improve the physics 
simulation:
\begin{itemize}
\item A new matrix element for single top production, including
both $s$- and $t$-channel mechanisms, has been implemented 
in the \HWPPClass{MEPP2SingleTop} class.
\item The parametrization of the kinematics for emissions in the dipole shower
   has been made more efficient.
\item Decays of $W$ and $Z$ bosons now use the POWHEG decayers by default.
\item The width treatment in BSM decay chains has been greatly improved
      and is now switched on by default in the \texttt{.model} files. To get the
      old behaviour, use
\begin{quote}\tt
     set /Herwig/NewPhysics/NewModel:WhichOffshell Selected
\end{quote}
\item The handling of beam remnants has been improved in multiple
   contexts, leading to a much lower error rate in hadronic collisions.
   An additional \HWPPParameterValueB{ClusterFissioner}{RemnantOption}{VeryHard}
   option for the  \HWPPParameter{ClusterFissioner}{RemnantOption}
   switch in the \HWPPClass{ClusterFissioner} class has been added
   to disable any special treatment of beam remnant
   clusters. The default remains unchanged.
\item The Higgs boson mass is now set to 125.9 GeV (from PDG 2013 update).
\end{itemize}

A number of technical changes have been made:
\begin{itemize}
\item To help with the coming transition to \texttt{C++-11}, we provide the new
      {\tt--enable-stdcxx11} configure flag. Please try to test builds with
   this flag enabled and let us know any problems, but do not use this
   in production code yet. In future releases, this flag will be on by
   default.
\item Many new Rivet analyses have been included in the \texttt{Tests} directory.
\item The header structure and compilation of the shower module
      has been improved. The parameters relating the evolution
      scale and Sudakov decomposition for radiation generated in 
      the shower have been grouped into one {\tt struct}.
\item The boolean flag describing whether or not POWHEG corrections
      were present in the \HWPPClass{HwMEBase} and \HWPPClass{HwDecayerBase}
      classes has been changed to an enumeration to allow better
      control over the types of emission.
\item  If any decay modes are selectively disabled, using the
       \HWPPClass{BranchingRatioReweighter} class as
       post-handler, {\it i.e.}
\begin{quote}\tt
create Herwig::BranchingRatioReweighter BRreweight

insert LHCGenerator:EventHandler:PostDecayHandlers 0 BRreweight
\end{quote}
 will cause all reported cross sections to include the
       branching ratio factor(s) from the previous stages correctly. Care should be
       taken not to use this option in processes, such as Higgs boson production in
       association with a $W^\pm$ or $Z^0$ boson, where the cross section already
       includes the branching ratio factor.
\item The search path for repository \texttt{read} command is now configurable on the
      command line with the \texttt{-i} and \texttt{-I} switches. By default, the
      installation location is now included in the search path, so that
\begin{quote}\tt
Herwig++ read LEP.in
\end{quote}
 will work in an empty directory. The current
   working directory will always be searched first. 
   The rarely used \texttt{Herwig++ init} command has been made consistent
   with \texttt{read} and \texttt{run} and should now be used without the \texttt{-i} flag.
\item Support for setting quark masses different from the values in the \ThePEGClass{ParticleData}
      objects in the \HWPPClass{O2AlphaS} class implementing the strong coupling has been enabled.
\item The various switches to turn off the compilation of BSM models have
      been unified into a single \texttt{--disable-models} switch. A new flag
      \texttt{--disable-dipole} can be used to turn off the compilation of the
      Dipole shower and Matchbox modules.
\item It is now possible to use anomalous vertices with the  \HWPPClass{MEfftoVH} matrix-element class.
\item A \ThePEGClass{Matcher} object has been added for charged, rather than all, leptons.
\item New scale choices have been added in the \HWPPClass{GeneralHardME} class to support testing
      against \textsf{Madgraph} results.
\item A number of changes have been made to the automatic calculation of running width effects in BSM models
      to improve the limits on the offshellness of the particles, speed and numerical stability
\item The consistency checks performed on SLHA and Les Houches event files have been significantly improved to reduce cases of GIGO.
\item The option of only including two-body decay modes in the running width calculations to speed up the calculation
       has been added.
\end{itemize}

The following bugs have been fixed:
\begin{itemize}
\item Various array bounds errors have been corrected.
\item Issues with inconsistent \texttt{C++} and \texttt{FORTRAN} compilers using OS X have been improved.
\item Vertex positions involving pseudo-vertices, that \HWPP\ inserts for technical reasons, are now set correctly.
\item The scale settings for MPI and the parton shower in POWHEG events has been changed to 
      fix reported anomalies in jet rates in POWHEG processes.
      NLO PDFs are now also set consistently in the example input files.
\item The interactive shell no longer quits following an error.
\item Possible division by zero errors in BSM branching ratio calculations have been fixed.
\item The limits on the momentum fraction allowed in forced splittings in the remnant handling have been
      fixed.
\item The colour rearrangement of beam clusters with each other is no longer allowed.
\item Various fixes have been made to ensure that the labelling of remnant clusters is correct
      after colour rearrangement and additional soft scatters.
\item Changes have been made to improve checkpointing in the decayer and matrix-element classes. The
      regular dump files are now consistent.
\item A bug affecting the lifetime and vertex positions in the decay $\pi^0\to e^+e^-\gamma$
   has been fixed. The virtual photon is no longer included in the event record.
\item A bug, introduced in Herwig++ 2.6.0,  in the diffraction code which
   would abort any runs has been fixed.
\item A bug in showering  from $gg \to h$ in the dipole shower has been fixed.
\item Various minor inconsistencies in the general hard matrix elements and decayers, which did not effect the built-in models, 
      have been fixed in order to support models generated from the UFO format.
\item A number of code features which caused warnings with recent compilers have been fixed.
\item A number of minor fixes to the SUSY couplings have been made to improve the consistency and handling of running masses effects.
\item Various issues with colour flows in BSM models introduced when sextet particles were added have been fixed.
\item The generation of QED radiation from massless charged particles coming from Les Houches event files has been
      forbidden to avoid numerical stability problems.
\end{itemize}

\section{Summary}

  \HWPP\,2.7 is the ninth version of the \HWPP\ program with a complete simulation of 
  hadron-hadron physics and contains a number of important improvements
  with respect to the previous
  version. The program has been extensively tested against
  a large number of observables from LHC, LEP, Tevatron and B factories.
  All the features needed for realistic studies for 
  hadron-hadron collisions are present. As always, we look forward to 
  feedback and input from users, especially from the Tevatron and LHC experiments.

  Our next major milestone is the release of version 3.0, which will be at least as
  complete as \HW\ in all aspects of LHC and linear-collider simulation.
%  Following the release of \HWPP\,3.0, we expect that support for the 
%  {\sf FORTRAN} program will cease.
%%%In practice it has already ceased!

\section*{Acknowledgements} 

This work was supported by Science and Technology Facilities Council
and the FP7 
Marie Curie Initial Training Network MCnetITN under
contract PITN-GA-2012-315877. JB, SG, SP, CR, TS, AS and 
BZ acknowledge 
support from the Helmholtz Alliance ``Physics at the Terascale''. AP
is supported in part by the Swiss National Science Foundation (SNF)
under contract 200020-149517 and by the European Commission through
the``LHCPhenoNet" Initial Training Network PITN-GA-2010-264564.  We
would like to thank all those who have reported issues with the
previous release.
  
\bibliography{Herwig++}

\providecommand{\href}[2]{#2}\begingroup\raggedright\begin{thebibliography}{10}

\bibitem{Bahr:2008pv}
M.~B\mbox{\"{a}}hr {\em et.~al.}, {\it {Herwig++ Physics and Manual}},  {\em
  Eur. Phys. J.} {\bf C58} (2008) 639--707,
  [\href{http://xxx.lanl.gov/abs/0803.0883}{{\tt arXiv:0803.0883}}].

\bibitem{UpdatedManual}
M.~B\mbox{\"{a}}hr {\em et.~al.} arXiv version (v3) of Ref.~\cite{Bahr:2008pv},
  updated on 02/12/2008.

\bibitem{Bahr:2008tx}
M.~B\mbox{\"{a}}hr {\em et.~al.}, {\it {Herwig++ 2.2 Release Note}},
  \href{http://xxx.lanl.gov/abs/0804.3053}{{\tt arXiv:0804.3053}}.

\bibitem{Bahr:2008tf}
M.~B\mbox{\"{a}}hr {\em et.~al.}, {\it {Herwig++ 2.3 Release Note}},
  \href{http://xxx.lanl.gov/abs/0812.0529}{{\tt arXiv:0812.0529}}.

\bibitem{Arnold:2012fq}
K.~Arnold, L.~d'Errico, S.~Gieseke, D.~Grellscheid, K.~Hamilton, {\em et.~al.},
  {\it {Herwig++ 2.6 Release Note}},
  \href{http://xxx.lanl.gov/abs/1205.4902}{{\tt arXiv:1205.4902}}.

\bibitem{Degrande:2011ua}
C.~Degrande, C.~Duhr, B.~Fuks, D.~Grellscheid, O.~Mattelaer, {\em et.~al.},
  {\it {UFO - The Universal FeynRules Output}},  {\em Comput.Phys.Commun.} {\bf
  183} (2012) 1201--1214, [\href{http://xxx.lanl.gov/abs/1108.2040}{{\tt
  arXiv:1108.2040}}].

\bibitem{Christensen:2008py}
N.~D. Christensen and C.~Duhr, {\it {FeynRules - Feynman rules made easy}},
  {\em Comput.Phys.Commun.} {\bf 180} (2009) 1614--1641,
  [\href{http://xxx.lanl.gov/abs/0806.4194}{{\tt arXiv:0806.4194}}].

\bibitem{Alloul:2013bka}
A.~Alloul, N.~D. Christensen, C.~Degrande, C.~Duhr, and B.~Fuks, {\it
  {FeynRules 2.0 - A complete toolbox for tree-level phenomenology}},
  \href{http://xxx.lanl.gov/abs/1310.1921}{{\tt arXiv:1310.1921}}.

\bibitem{Staub:2008uz}
F.~Staub, {\it {SARAH}},  \href{http://xxx.lanl.gov/abs/0806.0538}{{\tt
  arXiv:0806.0538}}.

\bibitem{Staub:2012pb}
F.~Staub, {\it {SARAH 3.2: Dirac Gauginos, UFO output, and more}},  {\em
  Computer Physics Communications} {\bf 184} (2013) pp. 1792--1809,
  [\href{http://xxx.lanl.gov/abs/1207.0906}{{\tt arXiv:1207.0906}}].

\bibitem{Staub:2013tta}
F.~Staub, {\it {SARAH 4: A tool for (not only SUSY) model builders}},
  \href{http://xxx.lanl.gov/abs/1309.7223}{{\tt arXiv:1309.7223}}.

\bibitem{Nason:2004rx}
P.~Nason, {\it A new method for combining {NLO} {QCD} with shower {M}onte
  {C}arlo algorithms},  {\em JHEP} {\bf 11} (2004) 040,
  [\href{http://xxx.lanl.gov/abs/hep-ph/0409146}{{\tt hep-ph/0409146}}].

\bibitem{Frixione:2007vw}
S.~Frixione, P.~Nason, and C.~Oleari, {\it {Matching NLO QCD computations with
  Parton Shower simulations: the POWHEG method}},  {\em JHEP} {\bf 11} (2007)
  070, [\href{http://xxx.lanl.gov/abs/0709.2092}{{\tt arXiv:0709.2092}}].

\bibitem{Richardson:2013jca}
P.~Richardson and A.~Wilcock, {\it {Monte Carlo Simulation of Hard Radiation in
  Decays in Beyond the Standard Model Physics in Herwig++}},
  \href{http://xxx.lanl.gov/abs/1303.4563}{{\tt arXiv:1303.4563}}.

\bibitem{Catani:2002hc}
S.~Catani, S.~Dittmaier, M.~H. Seymour, and Z.~Trocsanyi, {\it {T}he {D}ipole
  {F}ormalism for {N}ext-to-{L}eading {O}rder {QCD} {C}alculations with
  {M}assive {P}artons},  {\em Nucl. Phys.} {\bf B627} (2002) 189--265,
  [\href{http://xxx.lanl.gov/abs/hep-ph/0201036}{{\tt hep-ph/0201036}}].

\bibitem{Catani:1996vz}
S.~Catani and M.~Seymour, {\it {A General algorithm for calculating jet
  cross-sections in NLO QCD}},  {\em Nucl.Phys.} {\bf B485} (1997) 291--419,
  [\href{http://xxx.lanl.gov/abs/hep-ph/9605323}{{\tt hep-ph/9605323}}].

\bibitem{Alioli:2013nda}
S.~Alioli, S.~Badger, J.~Bellm, B.~Biedermann, F.~Boudjema, {\em et.~al.}, {\it
  {Update of the Binoth Les Houches Accord for a standard interface between
  Monte Carlo tools and one-loop programs}},
  \href{http://xxx.lanl.gov/abs/1308.3462}{{\tt arXiv:1308.3462}}.

\bibitem{Platzer:2011bc}
S.~Pl\mbox{\"{a}}tzer and S.~Gieseke, {\it {Dipole Showers and Automated NLO
  Matching in Herwig++}},  \href{http://xxx.lanl.gov/abs/1109.6256}{{\tt
  arXiv:1109.6256}}.

\bibitem{Platzer:Matching}
S.~Platzer, ``{Subleading effects in NLO plus parton shower matching, in
  preparation}.''.

\bibitem{Campanario:2013fsa}
F.~Campanario, T.~Figy, S.~Pl\mbox{\"{a}}tzer, and M.~Sj\mbox{\"{o}}dahl, {\it
  {Electroweak Higgs plus Three Jet Production at NLO QCD}},
  \href{http://xxx.lanl.gov/abs/1308.2932}{{\tt arXiv:1308.2932}}.

\bibitem{Plehn:1996wb}
T.~Plehn, M.~Spira, and P.~Zerwas, {\it {Pair production of neutral Higgs
  particles in gluon-gluon collisions}},  {\em Nucl.Phys.} {\bf B479} (1996)
  46--64, [\href{http://xxx.lanl.gov/abs/hep-ph/9603205}{{\tt
  hep-ph/9603205}}].

\bibitem{Dawson:1998py}
S.~Dawson, S.~Dittmaier, and M.~Spira, {\it {Neutral Higgs boson pair
  production at hadron colliders: QCD corrections}},  {\em Phys.Rev.} {\bf D58}
  (1998) 115012, [\href{http://xxx.lanl.gov/abs/hep-ph/9805244}{{\tt
  hep-ph/9805244}}].

\bibitem{Baur:2002qd}
U.~Baur, T.~Plehn, and D.~L. Rainwater, {\it {Determining the Higgs boson
  selfcoupling at hadron colliders}},  {\em Phys.Rev.} {\bf D67} (2003) 033003,
  [\href{http://xxx.lanl.gov/abs/hep-ph/0211224}{{\tt hep-ph/0211224}}].

\bibitem{Baur:2003gp}
U.~Baur, T.~Plehn, and D.~L. Rainwater, {\it {Probing the Higgs selfcoupling at
  hadron colliders using rare decays}},  {\em Phys.Rev.} {\bf D69} (2004)
  053004, [\href{http://xxx.lanl.gov/abs/hep-ph/0310056}{{\tt
  hep-ph/0310056}}].

\bibitem{Dolan:2012rv}
M.~J. Dolan, C.~Englert, and M.~Spannowsky, {\it {Higgs self-coupling
  measurements at the LHC}},  {\em JHEP} {\bf 1210} (2012) 112,
  [\href{http://xxx.lanl.gov/abs/1206.5001}{{\tt arXiv:1206.5001}}].

\bibitem{Papaefstathiou:2012qe}
A.~Papaefstathiou, L.~L. Yang, and J.~Zurita, {\it {Higgs boson pair production
  at the LHC in the $b \bar{b} W^+ W^-$ channel}},  {\em Phys.Rev.} {\bf D87}
  (2013) 011301, [\href{http://xxx.lanl.gov/abs/1209.1489}{{\tt
  arXiv:1209.1489}}].

\bibitem{Goertz:2013kp}
F.~Goertz, A.~Papaefstathiou, L.~L. Yang, and J.~Zurita, {\it {Higgs Boson
  self-coupling measurements using ratios of cross sections}},
  \href{http://xxx.lanl.gov/abs/1301.3492}{{\tt arXiv:1301.3492}}.

\bibitem{Goertz:2013eka}
F.~Goertz, A.~Papaefstathiou, L.~L. Yang, and J.~Zurita, {\it {Measuring the
  Higgs boson self-coupling at the LHC using ratios of cross sections}},
  \href{http://xxx.lanl.gov/abs/1309.3805}{{\tt arXiv:1309.3805}}.

\bibitem{Barr:2013tda}
A.~J. Barr, M.~J. Dolan, C.~Englert, and M.~Spannowsky, {\it {Di-Higgs final
  states augMT2ed -- selecting $hh$ events at the high luminosity LHC}},
  \href{http://xxx.lanl.gov/abs/1309.6318}{{\tt arXiv:1309.6318}}.

\bibitem{Dolan:2013rja}
M.~J. Dolan, C.~Englert, N.~Greiner, and M.~Spannowsky, {\it {Further on up the
  road: $hhjj$ production at the LHC}},
  \href{http://xxx.lanl.gov/abs/1310.1084}{{\tt arXiv:1310.1084}}.

\bibitem{deFlorian:2013jea}
D.~de~Florian and J.~Mazzitelli, {\it {Higgs pair production at NNLO}},
  \href{http://xxx.lanl.gov/abs/1309.6594}{{\tt arXiv:1309.6594}}.

\bibitem{Djouadi:1999rca}
A.~Djouadi, W.~Kilian, M.~Muhlleitner, and P.~Zerwas, {\it {Production of
  neutral Higgs boson pairs at LHC}},  {\em Eur.Phys.J.} {\bf C10} (1999)
  45--49, [\href{http://xxx.lanl.gov/abs/hep-ph/9904287}{{\tt
  hep-ph/9904287}}].

\bibitem{Gianotti:2002xx}
F.~Gianotti, M.~Mangano, T.~Virdee, S.~Abdullin, G.~Azuelos, {\em et.~al.},
  {\it {Physics potential and experimental challenges of the LHC luminosity
  upgrade}},  {\em Eur.Phys.J.} {\bf C39} (2005) 293--333,
  [\href{http://xxx.lanl.gov/abs/hep-ph/0204087}{{\tt hep-ph/0204087}}].

\bibitem{Moretti:2004wa}
M.~Moretti, S.~Moretti, F.~Piccinini, R.~Pittau, and A.~Polosa, {\it {Higgs
  boson self-couplings at the LHC as a probe of extended Higgs sectors}},  {\em
  JHEP} {\bf 0502} (2005) 024,
  [\href{http://xxx.lanl.gov/abs/hep-ph/0410334}{{\tt hep-ph/0410334}}].

\bibitem{Baglio:2012np}
J.~Baglio, A.~Djouadi, R.~Grober, M.~Muhlleitner, J.~Quevillon, {\em et.~al.},
  {\it {The measurement of the Higgs self-coupling at the LHC: theoretical
  status}},  \href{http://xxx.lanl.gov/abs/1212.5581}{{\tt arXiv:1212.5581}}.

\bibitem{Glover:1987nx}
E.~N. Glover and J.~van~der Bij, {\it {HIGGS BOSON PAIR PRODUCTION VIA GLUON
  FUSION}},  {\em Nucl.Phys.} {\bf B309} (1988) 282.

\bibitem{hpair}
``hpair program.'' {http://people.web.psi.ch/spira/hpair/}.

\bibitem{Alwall:2011uj}
J.~Alwall, M.~Herquet, F.~Maltoni, O.~Mattelaer, and T.~Stelzer, {\it {MadGraph
  5 : Going Beyond}},  {\em JHEP} {\bf 1106} (2011) 128,
  [\href{http://xxx.lanl.gov/abs/1106.0522}{{\tt arXiv:1106.0522}}].

\bibitem{rikkert}
R.~Frederix. Private Communication.

\bibitem{joselilin}
J.~Zurita and L.~L. Yang. Private Communication.

\bibitem{Abe:1997xk}
{\bf CDF} Collaboration, F.~Abe {\em et.~al.}, {\it {Double parton scattering
  in $\bar{p}p$ collisions at $\sqrt{s} = 1.8 $TeV}},  {\em Phys.Rev.} {\bf
  D56} (1997) 3811--3832.

\bibitem{Abazov:2009gc}
{\bf D0} Collaboration, V.~M. Abazov {\em et.~al.}, {\it {Double parton
  interactions in photon+3 jet events in $p p$ bar collisions $\sqrt{s}=1.96$
  TeV}},  {\em Phys.Rev.} {\bf D81} (2010) 052012,
  [\href{http://xxx.lanl.gov/abs/0912.5104}{{\tt arXiv:0912.5104}}].

\bibitem{Bahr:2013gkj}
M.~B{\"a}hr, M.~Myska, M.~H. Seymour, and A.~Si{\'o}dmok, {\it {Extracting
  $\sigma_{\!\textit{effective}}$ from the CDF gamma+3jets measurement}},  {\em
  JHEP} {\bf 1303} (2013) 129, [\href{http://xxx.lanl.gov/abs/1302.4325}{{\tt
  arXiv:1302.4325}}].

\bibitem{Abramowicz:2013iva}
H.~Abramowicz, P.~Bartalini, M.~B{\"a}hr, N.~Cartiglia, E.~Dobson, {\em
  et.~al.}, {\it {Summary of the Workshop on Multi-Parton Interactions (MPI@LHC
  2012)}},  \href{http://xxx.lanl.gov/abs/1306.5413}{{\tt arXiv:1306.5413}}.

\bibitem{Seymour:2013qka}
M.~H. Seymour and A.~Siodmok, {\it {Constraining MPI models using
  $\sigma_{\!\textit{effective}}$ and recent Tevatron and LHC Underlying Event
  data}},  \href{http://xxx.lanl.gov/abs/1307.5015}{{\tt arXiv:1307.5015}}.

\bibitem{CDFUEscan}
{\bf CDF} Collaboration, R.~Field, C.~Group, and D.~Wilson, {\it {The Energy
  Dependence of the Underlying Event in Hadronic Collisions}}, . talk presented
  by R. Field at the 36th International Conference on High Energy Physics
  (ICHEP 2012), Melbourne, Australia, July~5, 2012; and CDF physics note
  CDF/ANAL/CDF/CDFR/10874, unpublished.

\bibitem{Aad:2010fh}
{\bf ATLAS} Collaboration, G.~Aad {\em et.~al.}, {\it {Measurement of
  underlying event characteristics using charged particles in pp collisions at
  $\sqrt{s} = 900 GeV$ and 7 TeV with the ATLAS detector}},  {\em Phys. Rev. D}
  {\bf 83} (2011) 112001, [\href{http://xxx.lanl.gov/abs/1012.0791}{{\tt
  arXiv:1012.0791}}].

\end{thebibliography}\endgroup
\end{document}